\documentstyle[12pt]{article}
\catcode `\@=11                                            
\@addtoreset{equation}{section}                            
   
\catcode `\@=12                                            
\newcommand\jjed{{\mathchoice {\rm 1\mskip-4mu l} {\rm 1\mskip-4mu l}  
{\rm 1\mskip-4.5mu l} {\rm 1\mskip-5mu l}}}                     
\begin{document}

\title{To what extent do the Classical Equations of Motion Determine
the Quantization Scheme? }
\author{J. Cis{\l}o, J. {\L}opusza\'{n}ski \\
    Institute of Theoretical Physics \\
         University of Wroc{\l}aw \\
pl. M. Borna 9, 50-205 Wroc\l aw, Poland}
\maketitle

\begin{abstract}
A simple example of one particle moving in a (1+1) space-time is
considered. As an example we take the harmonic oscillator.
We confirm the statement that the classical Equations of Motion
do not determine at all the quantization scheme. To this aim we use two
inequivalent Lagrange functions, yielding Euler-Lagrange Equations, having
the same set of solutions. We present in detail the calculations of both
cases to emphasize the differences occuring between them.
\end{abstract}


\section{Introduction}
In classical physics the content of dynamical process is mainly
characterized by the Equations of Motion of the physical system.
It may happen that these equations can be derived from a certain Lagrange
function as the Euler-Lagrange-Equations. If this is the case we may expect
that there can be many (even infinitely many!) nonequivalent Lagrange
functions linked to these equations of motion, yielding the same set
of solutions -- so called s-equivalent equations. These fact is well known
[1,2].

Two nonequivalent but s-equivalent Lagrange functions lead to two distinct
Hamilton functions and distinct canonical momentum variables. If we take
as the starting point for our consideration these two Hamilton functions
as well as these two sets of canonical momenta and try to quantize them in
a standard way we get, in general, two distinct quantization schemes,
differing essentially from each other, althought both having common
roots coming from the same equations of motion. This fact is not new and
well known to some physicists [1,2], but -- strange enough -- not much
attention was paid by them to this problem.

From what was said so far we may infer that the answer to the question
posed in the title is that the equations of motion do not determine
the quantization scheme.

Below we shall present an elucidating example in favour of the statement
made above.

One remark is here in order. If we choose two inequivalent, but
s-equivalent Lagrange functions, say, $L$, and $L'$, we get two sets
of canonical variables
\begin{equation}
\label{11}
(x,p \equiv {\partial L  \over \partial \dot{x} })
\quad \hbox{and} \quad
(x,p' \equiv {\partial L' \over \partial \dot{x} }). 
\end{equation}
Notice that those canonical variables are not connected to each other
by a canonical transformation, viz.
\begin{equation}
\label{12}
(x,p) \rightarrow (X(x,p),P(x,p)).
\end{equation}
Should they be linked to each other by a point transformation
\begin{equation}
\label{13}
(x,p) \rightarrow (x,p'(x,p))
\end{equation}
it would follow from the canonical Poisson brackets that
\begin{equation}
\label{14}
p'=p+f(x), \quad f(x)- \hbox{arbitrary function},
\end{equation}
which is not the case considered by us in this note.


\section{The case of one classical particle in a (1+1) space-time.
The Master Equation for $H'(x,p')$}
To buttress the above observations and make plain the goal of this note,
we shall investigate a very simple problem of classical mechanics.
So let us restrict ourselves to the case of one classical particle,
moving in a (1+1) space-time. Let us take sufficiently smooth equation
of motion
\begin{equation}
\label{21}
   \ddot{x}=f(x,\dot{x},t)
\end{equation}
where $x(t)$ denotes the location of the particle and
$\dot{x}\equiv dx / dt$ and $\ddot{x}\equiv d^{2}x / dt^{2}$ denote its
velocity and acceleration resp., $t$ being the independent time variable.
To every such equation belongs a Lagrange function [2,3],
$L(x,\dot{x},t)$. Let us further restrict ourselves for simplicity reason
to automonous Lagrange functions. It is known [2,4] that the most general
expression for Lagrange function, $L'(x,\dot{x})$, s-equivalent to
$L(x,\dot{x})$, is
\begin{equation}
\label{22}
   L'=\dot{x} \int\limits^{\dot{x}}_{c} {d \Sigma (H) \over dH }
   {\partial^{2} L \over \partial \dot{x}^{2} } |_{\dot{x}=u} du
   - \Sigma(H)
\end{equation}
where $\Sigma(H)$ is an arbitrary differentiable function of $H$,
different from zero a.e., and
\begin{equation}
\label{23}
    H\equiv\dot{x}{\partial L \over \partial \dot{x} } - L,
\end{equation}
the Hamilton function. The constant $c$ is so choosen that the integral
on the r.h.s. of (\ref{22}) does not diverge. It is easy to see that
we have
\footnote{ The presence of $d \Sigma(H) / dH$ in (\ref{24}) causes that
the set of solutions of equation on the l.h.s. of (\ref{24}) and the one
on the r.h.s. differ by a set of measure zero.}
\begin{equation}
\label{24}
    {\partial L' \over \partial x}- \dot{x}
    {\partial^{2} L' \over \partial \dot{x} \partial x }
    -\ddot{x} {\partial^{2} L' \over \partial \dot{x}^{2}}=
    { d\Sigma(H) \over dH} \left(
    {\partial L \over \partial x}- \dot{x}
    {\partial^{2} L \over \partial \dot{x} \partial x }
    -\ddot{x} {\partial^{2} L \over \partial \dot{x}^{2}} \right).
\end{equation}
Thus $L'$ is, indeed, s-equivalent to $L$. We have
\begin{equation}
\label{25}
    H' \equiv \dot{x} {\partial L' \over \partial \dot{x} } - L' =
    \Sigma(H).
\end{equation}
We get also
\begin{equation}
\label{26}
    p' \equiv {\partial L' \over \partial \dot{x}}.
\end{equation}
It is trivial to find $H'$ and $p'$ for given $\Sigma$ and $L$ as functions
of $x$ and $\dot{x}$. It is, however, not so simple to get $H'$ as a
function of $x$ and $p'$. To get that let as observe that
from (\ref{26}) and (\ref{22}) follows
\begin{equation}
\label{27}
    1={ d\Sigma \over dH} {\partial^{2}L \over \partial \dot{x}^{2} }
    {\partial \dot{x}(x,p') \over \partial p'}.
\end{equation}
Taking into account the relation
\begin{equation}
\label{28}
   {\partial H'(x,p') \over \partial p'} = \dot{x}(x,p')
\end{equation}
we get from (\ref{27}) and (\ref {28})
\begin{equation}
\label{29}
   {\partial^{2} H'(x,p') \over \partial p'^{2}} =
   {\partial \dot{x}(x,p') \over \partial p'} =
   \left( {dH' \over dH} {\partial^{2}L \over \partial \dot{x}^{2} }
   |_{\dot{x}=\partial H' / \partial p'} \right) ^{-1}.
\end{equation}
This is the Master Equation for $H'$ as a function of $p'$ and $x$; it is,
in general, nonlinear.

The solutions of this Master Equation have to satisfy a physically justified
requirement that $p'$ has to tend to zero as sun as $\dot{x}$ tends to
zero and vice versa, or in other words
\begin{equation}
\label{210}
    {\partial H' \over \partial p'}|_{p'=0} =0.
\end{equation}


\section{Application of the Master Equation}
To make use of this Master Equation one has, of course, to specify what $L$
and $\Sigma$ are. This will be done now. We choose
\begin{equation}
\label{31}
   L={1 \over 2} \dot{x}^{2} - V(x)
\end{equation}
where $V(x)$ is a nonnegative function of $x$
\footnote{ In case $V(x)$ is just bounded from below we may make it
nonnegative for each $x$ by adding to it a suitably choosen positive
constant.}
and
\begin{equation}
\label{32}
   \Sigma(H) = H' = \sqrt{2 H}
\end{equation}
Square root means nonnegative root.
For this choice of $L$ and $\Sigma$
\footnote{ The equality (\ref{32}) should be understood as follows
   $$ H'= \alpha \sqrt{2H} , \quad \alpha-\hbox{constant} $$
As
$H'$ as well
$H$ should have the same dimensions it follows that
the constant
$\alpha$ has to have the dimension
   $$ [\alpha]= g^{1/2} cm /sec $$
The equation (\ref{33}) reads
  $${\partial^{2} H' \over \partial p^{2} }={1 \over \alpha^{2} m } H' $$
as
$L$ in (\ref{31}) becomes
   $$ L= {1 \over 2} m \dot{x}^{2} - V(x) $$
where $m$ denotes the mass of the particle. In this note we put
   $$ \alpha=m=1. $$
   }
equation (\ref{29}) reduces to
\begin{equation}
\label{33}
   {\partial^{2} H' \over \partial p'^{2} } = H'.
\end{equation}
The solution of (\ref{33}) reads
\begin{equation}
\label{35}
    H'=\alpha(x) \sinh p' + \beta(x) \cosh p'.
\end{equation}
This solution has to satisfy the requirement (\ref{210}) and therefore
\begin{equation}
\label{alpha}
   \alpha(x)=0.
\end{equation}
From (\ref{31}) and the Hamilton Equation, we obtain
\begin{equation}
\label{39}
     {d \over dt} {\partial H' \over \partial p' } =\ddot{x}=-
     {\partial V \over \partial x}.
\end{equation}
If we insert in (\ref{39})
\begin{equation}
\label{310}
    \dot{x}=\beta(x) \sinh  p'
\end{equation}
\begin{equation}
\label{311}
    \dot{p'}=-{\partial H' \over \partial x} =
    - {\partial \beta \over \partial x } \cosh p',
\end{equation}
following from the Hamilton Equations, we get
$$
   {1 \over 2} {\partial \over \partial x} \beta^{2} =
   {\partial V \over \partial x}
$$
or
\begin{equation}
\label{312}
    \beta=\pm \sqrt{2(V+C)}
\end{equation}
$C$ being a constant. Hence
\begin{equation}
\label{313}
     H'=\pm \sqrt{2(V(x)+C)} \cosh p'.
\end{equation}
According to our choice (\ref{32}) taking into account (\ref{312})
we should have
\begin{equation}
\label{314}
  \pm \sqrt{2(V+C)} \cosh p' = \sqrt{ 2(\dot{x}^{2}/2+V)}
\end{equation}
To keep both sides of the relation (\ref{314}) compatible with each other
we have to choose the (+) sign on the l.h.s. of (\ref{314}). Since for $p'$
tending to zero $\dot{x}$ should also tend to zero we conclude that $C=0$.
Thus eventually we have
\begin{equation}
\label{316}
    H'= \sqrt{2 V(x)}\cosh p'.
\end{equation}

Notice that we could  as well choose 
in the definition on the r.h.s. of (\ref{32}) the (--) sign in front
of the root or use both signs suitable for certain nonoverlapping intervals
of the variable $x$. This would cause a change of our model. 
In each case, mentioned above,
the Hamilton Equations are s-equivalent to original equations of motions
and the Hamilton functions are constants of motion.

The original Hamilton function reads
\begin{equation}
\label{319}
    H={1 \over 2} p^{2} + V(x).
\end{equation}


\section{Example of the harmonic oscillator}
For the case of the harmonic oscillator
\begin{equation}
\label{320}
   V(x)={1 \over 2} x^{2}
\end{equation}
and we choose the model
%
%
\footnote{ Relation (\ref{321}) has to be understood as follows
$$
    H'=\sqrt{m} \omega \alpha x  \cosh
    \left(    {p' \over \sqrt{m} \alpha}  \right)
$$
    as we have
$
   L={1 \over 2} m \dot{x}^{2}-{1\over 2} m \omega^{2} x^{2},
   \quad  [\omega]=1/sec. \quad
$
 We put $ m = \alpha = \omega =1 $. }
\begin{equation}
\label{321}
   H'= x \cosh p'.
\end{equation}
$H'$ is not bounded from below. As $H'$ is a conserved quantity the
singularity of (\ref{321}) appears at $x=0$. It can be easily removed
by taking as the potential
\begin{equation}
\label{322}
    V(x)={1 \over 2} x^{2} + a,
\end{equation}
$a$ an arbitrary positive constant. We are not going to use this procedure
as it would complicate essentially our further calculations. For the case
(\ref{321}) the classical trajectories would be given by
\begin{equation}
\label{323}
    x \cosh p'= b, \quad b-\hbox{a constant}.
\end{equation}
For $b>0$ the phase trajectory lies in the strip $0<x\leq b$ for $b<0$,
in $b\leq x<0$, while $-\infty <p'< +\infty $.
If $x \rightarrow b$ then  $ dp' / dx \rightarrow \infty$.

It is easy to see that (\ref{321}) gives rise to the equation
\begin{equation}
\label{324}
    \ddot{x}+x=0.
\end{equation}
Indeed, we have
\begin{equation}
\label{325}
    {\partial H' \over \partial x} = \cosh p' =  - \dot{p}', \quad 
    {\partial H' \over \partial p'}= x \sinh p' = \dot{x}.
\end{equation}
Then
\begin{equation}
\label{326}
    \ddot{x}=\dot{x} \sinh p' + \dot{p'}x \cosh p' =
    x (\sinh p')^{2} - x(\cosh p' )^{2} = -x.
\end{equation}


\section{Quantization}
Let us now try to quantize $H$ and $H'$. We assume that the operators
$(x,\hat{p})$ and $(x,\hat{p}')$ satisfy the standard canonical
commutation relations. Then $\hat{p}$ as well as $\hat{p}'$ can be
replaced by
\begin{equation}
\label{41}
   -i \partial_{x}
\end{equation}
in formulae (\ref{319}) and (\ref{316}) resp. We get the following
differential expressions, viz.
\begin{equation}
\label{42}
   H_{Q}=-{1 \over 2} (\partial_{x})^{2}+V(x)
\end{equation}
\begin{equation}
\label{43}
   H_{Q}'=\sqrt{{1 \over 2} V(x)} \cos (\partial_{x}) +
          \cos (\partial_{x} ) \sqrt{ {1 \over 2} V(x)}
\end{equation}
This expressions applied to $C^{\infty}_{0}({\bf R})$ define symmetric
operators in $L^{2}({\bf R})$
\footnote{As it is well known $H_{Q}$ can be extended to a self-adjoint
operator}.
Notice that the operator (\ref{43}) is not local, viz.
\begin{equation}
\label{45}
   \cos ( \partial_{x} ) \Psi(x)=
   {1 \over 2} \left( \Psi(x+i) + \Psi(x-i) \right)
\end{equation}
and therefore (we denote the operator also by $H'_{Q}$)
\begin{equation}
\label{45a}
    H'_{Q} \Psi(x) =
  {1 \over \sqrt{2} } \left( \sqrt{V(x)}+\sqrt{V(x+i)} \right) \Psi(x+i) +
\end{equation}
$$
 +  {1 \over \sqrt{2} } \left( \sqrt{V(x)}+\sqrt{V(x-i)} \right) \Psi(x-i)    
$$


\section{Harmonic oscillator}
The potential is given by (\ref{320}). As it is well known the eigenvalues
for $H_{Q}$ are
\begin{equation}
\label{46}
    n+{1 \over 2}, \quad n-\hbox{natural number or  0}
\end{equation}
and the eigenfunctions are the Hermitean functions
\begin{equation}
\label{47}
   \Psi_{n}=exp(-x^{2}/2) H_{n}(x)
\end{equation}
where $H_{n}$ are the Hermitean polynomials [5].

For the case $H_{Q}'$ we have (see(\ref{43}))
\begin{equation}
\label{48}
   H_{Q}'={1 \over 2} (x \cos(\partial_{x}) + \cos(\partial_{x}) x).
\end{equation}
The case of nonlocal $H_{Q}'$ will be investigated in Section 7.

It seems more convenient to start the discussion by using different
canonically conjugate variables, namely (hereafter we shall use the letter
$p$ instead of $p'$)
\begin{equation}
\label{410a}
   -i\partial_{x} \rightarrow p \quad \hbox{and} \quad
   x \rightarrow i \partial_{p}.
\end{equation}
Notice that the two systems of variables are linked
by a Fourier transformation.

Let us denote the new Hamilton operator by $K$.
Then we get from (\ref{48})
\begin{equation}
\label{410b}
   K={i \over 2} (\partial_{p} \cosh(p) + \cosh(p) \partial_{p}).
\end{equation}

The differential expression (\ref{410b}) when applied to
$C_{0}^{\infty}({\bf R})$
defines a symmetric operator in $L^{2}({\bf R})$, which we shall also denote
by $K$. This statement as well as the following results are dicussed
{\it in extenso} in the Appendix. It is shown that for real $\gamma$
\begin{equation}
\label{411}
   0 \leq \gamma < 2
\end{equation}
the system of functions
\begin{equation}
\label{412}
   \{\Psi_{2n+\gamma}(p)\}_{ n \in {\bf Z}},
\end{equation}
where
\begin{equation}
\label{413}
    \Psi_{2n+\gamma}(p)={ 1 \over \sqrt{ \pi \cosh p} } 
    \exp(-i (2n+\gamma) \arctan \sinh p)
\end{equation}
are the solution of the equations
\begin{equation}
\label{414}
    K \Psi_{2n+\gamma}(p) = (2n+\gamma) \Psi_{2n+\gamma}(p),
\end{equation}
is an orthonormal basis for $L^{2}({\bf R})$. Thus for each fixed $\gamma$
of the interval (\ref{411}) $K$ has a self-adjoint extension
$K_{\alpha}, \quad \alpha \equiv exp(-i\gamma \pi)$.


                                                                                                                                                               
\section{Fourier transform of the eigenfunctions}                                                                                                              
In this section we investigate Fourier transforms of eigenfuntions                                                                                             
of the Hamilton operator (\ref{410b}). We find an expression for                                                                                               
the Fourier transform in the case when                                                                                                                         
eigenvalues are equal to $n+1/2$, where $n$ is integer.                                                                                                        
However, we do not see how to solve the problem for other eigenvalues.                                                                                         
We show that in the considered case the Fourier transforms                                                                                                     
are eigenfuntions of nonlocal Hamilton operator (\ref{48}).                                                                                                    
These eigenfunctions form two bases in the Hilbert space.                                                                                                      
Additionally  we get a family of orthogonal polynomials with                                                                                                   
the weight $( \cosh \pi x )^{-1}$.

                                                                                                                                                               
Let us rewrite the result (\ref{413}). The normalized eigenfunction                                                                                            
belonging to the eigenvalue $\lambda$ reads                                                                                                                    
\begin{equation}                                                                                                                                               
\label{ef}                                                                                                                                                     
  \Psi_{\lambda}(p)={1 \over \sqrt{ \pi \cosh p} }                                                                                                             
  \exp (-i \lambda \arctan \sinh p).                                                                                                                           
\end{equation}                                                                                                                                                 
                                                                                                                                                               
We start with two identities:                                                                                                                                  
\begin{equation}                                                                                                                                               
\label{i1}                                                                                                                                                     
  {1 \over \sqrt{ \cosh p} }                                                                                                                                   
  \exp (- {i \over 2} \arctan \sinh p )                                                                                                                        
   = { 1+i \over 1+ie^{p} } e^{ p/2 },                                                                                                                         
\end{equation}                                                                                                                                                 
\begin{equation}                                                                                                                                               
\label{i2}                                                                                                                                                     
   \exp ( -i \arctan \sinh p ) = i { 1-ie^{p} \over 1+ie^{p} }.                                                                                                
\end{equation}                                                                                                                                                 
To check these identities it is most simple to compare the moduli and                                                                                          
arguments of the complex number on both sides of the equalities.                                                                                               
                                                                                                                                                               
Mulitiplying first identity by $n-th$ power of the second identity we                                                                                          
obtain                                                                                                                                                         
\begin{equation}                                                                                                                                               
\label{nef}                                                                                                                                                    
    \Psi_{n+ 1/2 } (p) = {1 \over \sqrt{\pi} }                                                                                                                 
    \left( i {1-ie^{p} \over 1+ie^{p} } \right)^{n}                                                                                                            
    { 1+i \over 1+ie^{p} } e^{p/2}.                                                                                                                            
\end{equation}                                                                                                                                                 
%
%
To get the Fourier transform of (\ref{nef})                                                                                                                    
\begin{equation}                                                                                                                                               
\label{dft}                                                                                                                                                    
    \Phi_{\lambda}(x) \equiv {1 \over \sqrt{2 \pi} }                                                                                                           
    \int\limits_{-\infty}^{+\infty} \Psi_{\lambda}(p) e^{ipx} dp                                                                                               
\end{equation}                                                                                                                                                 
we employ the method of generating function. We define                                                                                                         
\begin{equation}                                                                                                                                               
\label{gf}                                                                                                                                                     
    \Psi(p,t) \equiv \sum\limits_{n=0}^{\infty}                                                                                                                
    \Psi_{n+1/2}(p)t^{n} = {1 \over \sqrt{\pi} }                                                                                                               
    { 1+i \over (1-it)+i(1+it)e^{p} } e^{p/2}.                                                                                                                 
\end{equation}                                                                                                                                                 
The computation of (\ref{gf}) amounts to summing up the geometrical series                                                                                     
covergent for $|t|<1$. The Fourier transform of the generating function                                                                                        
(\ref{gf}) yields the generating function for the Fourier transforms                                                                                           
of the eigenfunctions:                                                                                                                                         
\begin{equation}                                                                                                                                               
\label{ftgf}                                                                                                                                                   
    \Phi(x,t) \equiv {1 \over \sqrt{2 \pi} }                                                                                                                   
    \int\limits_{-\infty}^{+\infty} \Psi(p,t) e^{ipx} dp                                                                                                       
    =\sum\limits_{n=0}^{\infty} \Phi_{n+1/2}(x) t^{n}.                                                                                                         
\end{equation}                                                                                                                                                 
                                                                                                                                                               
                                                                                                                                                               
To evaluate the integral (\ref{ftgf}) we shall use the method of complex                                                                                       
analysis. Let us consider the function                                                                                                                         
\begin{equation}                                                                                                                                               
\label{cf}                                                                                                                                                     
   {1 \over \sqrt{2 \pi} } \Psi(p,t) e^{ipx}                                                                                                                   
\end{equation}                                                                                                                                                 
as the function of a complex variable $p$ and let us compute the integral                                                                                      
of function (\ref{cf}) along the contour of the rectangle with the                                                                                             
vertices located at the points                                                                                                                                 
$(-a,0),\;(a,0),\;(a,2\pi i),\; (-a,2 \pi i),\; a>0$,                                                                                                          
running in the counterclockwise direction.                                                                                                                     
In the limit when $a$ tends to infinity, the integral along the lower side                                                                                     
yields $\Phi(x,t)$.                                                                                                                                            
To get the integral along the upper side of the rectangle                                                                                                      
we exploit the relation                                                                                                                                        
\begin{equation}                                                                                                                                               
\label{ii}                                                                                                                                                     
   {1 \over \sqrt{2 \pi} } \int\limits_{-\infty}^{+\infty}                                                                                                     
   \Psi(p+2 \pi i ,t) e^{i(p+2 \pi i)x} dp = - \Phi(x,t)e^{-2 \pi x}                                                                                           
\end{equation}                                                                                                                                                 
which follows from the property                                                                                                                                
\begin{equation}                                                                                                                                               
\label{2pi}                                                                                                                                                    
     \Psi(p+2\pi i,t) = - \Psi(p,t).                                                                                                                           
\end{equation}                                                                                                                                                 
In the limit the contributions from both remaining sides of the rectangle                                                                                      
vanish. Then the integral along the rectangle in the limit is equal to                                                                                         
\begin{equation}                                                                                                                                               
\label{il}                                                                                                                                                     
   \Phi(x,t) + \Phi(x,t) e^{-2\pi x}.                                                                                                                          
\end{equation}                                                                                                                                                 
Function (\ref{gf}) is a meromorphic function and has inside of the                                                                                            
rectangle a simple pole at the point                                                                                                                           
\begin{equation}                                                                                                                                               
\label{pr}                                                                                                                                                     
    \tilde{p}={i \pi \over 2} - 2 i \arctan t.                                                                                                                 
\end{equation}                                                                                                                                                 
                                                                                                                                                               
We can express the integral of the function (\ref{gf}) along the rectangle                                                                                     
by residuum of the function at the point $\tilde{p}$ which is equal to                                                                                         
\begin{equation}                                                                                                                                               
\label{res}                                                                                                                                                    
   - {i \over \pi} e^{- \pi x / 2}                                                                                                                             
   {1 \over \sqrt{1+t^{2}} }                                                                                                                                   
   \exp ( 2 x \arctan t ).                                                                                                                                     
\end{equation}                                                                                                                                                 
The integral (\ref{il}) is the product of $2 \pi i$ and                                                                                                        
the residuum (\ref{res}). Finally, we get                                                                                                                      
\begin{equation}                                                                                                                                               
\label{vgf}                                                                                                                                                    
    \Phi(x,t)={2 e^{-\pi x /2} \over 1 + e^{-2 \pi x} }                                                                                                        
    {1 \over \sqrt{1+t^{2}} }                                                                                                                                  
    \exp (2 x \arctan t ).                                                                                                                                     
\end{equation}                                                                                                                                                 
%
%
Let us set                                                                                                                                                     
\begin{equation}                                                                                                                                               
\label{pgf}                                                                                                                                                    
    W(x,t) \equiv { 1 \over \sqrt{1+t^{2}} }                                                                                                                   
    \exp ( 2 x \arctan t )                                                                                                                                     
    =\sum\limits_{n=0}^{\infty} W_{n}(x) {t^{n} \over n! }.                                                                                                    
\end{equation}                                                                                                                                                 
The formula (\ref{pgf}) defines the sequence of polynomials $W_{n}(x)$.                                                                                        
Degree of the polynomial $W_{n}(x)$ equals $n$.                                                                                                                
The polinomials $W_{n}(x)$                                                                                                                                     
are even functions for even $n$ and aod functions for add $n$.                                                                                                 
                                                                                                                                                               
Comparing definitions (\ref{ftgf}) and (\ref{pgf}) and the formula                                                                                             
(\ref{vgf}) we can write                                                                                                                                       
\begin{equation}                                                                                                                                               
\label{phin}                                                                                                                                                   
   \Phi_{n+1/2}(x)={2e^{-\pi x/2} \over 1+e^{-2\pi x} } {W_{n}(x) \over n!}                                                                                    
   =\Phi_{1/2}(x) {W_{n}(x) \over n!}.                                                                                                                         
\end{equation}                                                                                                                                                 
This formula holds for nonnegative integer $n$.                                                                                                                
From the definition of the Fourier transformation (\ref{dft}) and from the                                                                                     
relation                                                                                                                                                       
\begin{equation}                                                                                                                                               
\label{psiminus}                                                                                                                                               
   \Psi_{-\lambda}(p)=\overline{\Psi_{\lambda}(p)}                                                                                                             
\end{equation}                                                                                                                                                 
follows                                                                                                                                                        
\begin{equation}                                                                                                                                               
\label{phiminus}                                                                                                                                               
   \Phi_{-\lambda}(x)=\Phi_{\lambda}(-x).                                                                                                                      
\end{equation}                                                                                                                                                 
Thus for nonnegative integer $n$ we have                                                                                                                       
\begin{equation}                                                                                                                                               
\label{phinminus}                                                                                                                                              
   \Phi_{-n-1/2}(x)                                                                                                                                            
   =(-1)^{n}{2e^{\pi x/2} \over 1+e^{2\pi x} } {W_{n}(x) \over n!}                                                                                             
   =(-1)^{n}\Phi_{-1/2}(x) {W_{n}(x) \over n!}                                                                                                                 
\end{equation}                                                                                                                                                 
which supplements relation (\ref{phin}).                                                                                                                       
                                                                                                                                                               
                                                                                                                                                               
Let us investigate the polynomials $W_{n}(x)$. For this aim we                                                                                                 
differentiate$W(x,t)$, given by (\ref{pgf}), with respect to $t$. We get                                                                                       
\begin{equation}                                                                                                                                               
\label{dwdt}                                                                                                                                                   
   {\partial W(x,t) \over \partial t}                                                                                                                          
   = {2x-t \over 1+t^{2} } W(x,t).                                                                                                                             
\end{equation}                                                                                                                                                 
If we multiple both sides of (\ref{dwdt}) by $(1+t^{2})$ and compare the                                                                                       
coefficients of the same power of $t$ on both sides of (\ref{dwdt})                                                                                            
we obtain                                                                                                                                                      
\begin{equation}                                                                                                                                               
\label{w123}                                                                                                                                                   
\cases{                                                                                                                                                        
   W_{0}(x)=1 \cr                                                                                                                                              
   W_{1}(x)=2x \cr                                                                                                                                             
   W_{2}(x)=4x^{2}-1 \cr                                                                                                                                       
   W_{3}(x)=8x^{3}-10x \cr                                                                                                                                     
   W_{4}(x)=16x^{4}-56x^{2}+9 \quad \hbox{etc.}                                                                                                                
}                                                                                                                                                              
\end{equation}                                                                                                                                                 
as well as recurence formula                                                                                                                                   
\begin{equation}                                                                                                                                               
\label{wrec}                                                                                                                                                   
   W_{n+1}(x)+n^{2}W_{n-1}(x)=2xW_{n}(x), \quad n>0.                                                                                                           
\end{equation}                                                                                                                                                 
                                                                                                                                                               
                                                                                                                                                               
Now we are going to show that the Fourier transformed functions                                                                                                
$\Phi_{n+1/2}$ are eigenfunctions of the nonlocal Hamilton operator                                                                                            
(\ref{48}):                                                                                                                                                    
\begin{equation}                                                                                                                                               
\label{hphi}                                                                                                                                                   
    H_{Q}'\Phi_{n+1/2}(x) \equiv                                                                                                                               
    {i \over 2} \left( {1 \over 2} -ix \right) \Phi_{n+1/2}(x+i)-                                                                                              
    {i \over 2} \left( {1 \over 2} +ix \right) \Phi_{n+1/2}(x-i)                                                                                               
\end{equation}                                                                                                                                                 
$$                                                                                                                                                             
    = (n+1/2) \Phi_{n+1/2}(x).                                                                                                                                 
$$                                                                                                                                                             
We shall prove (\ref{hphi}) for nonnegative integer $n$; for negative                                                                                          
ones the proof is very similar to that for nonnegative.                                                                                                        
                                                                                                                                                               
Taking into account the relation                                                                                                                               
\begin{equation}                                                                                                                                               
\label{phi0pm}                                                                                                                                                 
    \Phi_{1/2}(x \pm i) = \mp i \Phi_{1/2}(x)                                                                                                                  
\end{equation}                                                                                                                                                 
and the formula (\ref{phin}) we conclude that (\ref{hphi}) holds iff                                                                                           
\begin{equation}                                                                                                                                               
\label{hwn}                                                                                                                                                    
h W_{n}(x) \equiv {1 \over 2} \left( {1 \over 2} -ix \right) W_{n}(x+i) +                                                                                      
                  {1 \over 2} \left( {1 \over 2} +ix \right) W_{n}(x-i) =                                                                                      
                      (n+1/2) W_{n}(x).                                                                                                                        
\end{equation}                                                                                                                                                 
To prove relation (\ref{hwn}) let us apply the expression $h$                                                                                                  
upon $W(x,t)$, namely                                                                                                                                          
\begin{equation}                                                                                                                                               
\label{hw}                                                                                                                                                     
   h W(x,t) = {1 \over 2} \left( {1 \over 2} -ix \right) W(x+i,t) +                                                                                            
              {1 \over 2} \left( {1 \over 2} +ix \right) W(x-i,t)                                                                                              
\end{equation}                                                                                                                                                 
$$                                                                                                                                                             
   =\left( {1 \over 2} + t {2x-t \over 1+t^{2} } \right) W(x,t) =                                                                                              
   {1 \over 2} W(x,t) + t { \partial W(x,t) \over \partial t}.                                                                                                 
$$                                                                                                                                                             
The last equality follows from the formula (\ref{dwdt}).                                                                                                       
If we compare the coefficients of the same power of $t$ on both                                                                                                
sides of (\ref{hw}) we get (\ref{hwn}). This completes the proof.                                                                                              
                                                                                                                                                               
                                                                                                                                                               
Let us return to the consideration of the previous section.                                                                                                    
There we learned that for any fixed $\lambda$ the functions                                                                                                    
$\Psi_{2n+\gamma }$, $n=0,\pm 1,\pm 2, \pm 3,...$,                                                                                                             
form an orthonormal basis in $L^{2}({\bf R})$.                                                                                                                 
It is known that the Fourier transformation maps an orthonormal                                                                                                
basis into a new orthonormal basis.                                                                                                                            
                                                                                                                                                               
We have found the Fourier transforms of the eigenfunctions only for                                                                                            
$\gamma =1/2$ and $\gamma = 3/2$. Further we choose $\gamma =1/2$.                                                                                             
and consider orthonormal basis:                                                                                                                                
\begin{equation}                                                                                                                                               
\label{basis}                                                                                                                                                  
   \Phi_{2n+1/2}, \quad n=0, \pm 1, \pm 2, \pm 3, ...                                                                                                          
\end{equation}                                                                                                                                                 
Taking into account the definition (\ref{phin}) and (\ref{phiminus})                                                                                           
and property of polynomials $W_{m}(-x)=(-1)^{m}W_{m}(x)$                                                                                                       
we get for nonnegative integer $n$  and $k$                                                                                                                    
\begin{equation}                                                                                                                                               
\label{sp1}                                                                                                                                                    
   \int\limits_{-\infty}^{+\infty}                                                                                                                             
   \Phi_{2n+1/2}(x) \Phi_{2k+1/2}(x) dx=                                                                                                                       
\end{equation}                                                                                                                                                 
$$                                                                                                                                                             
  = \int\limits_{-\infty}^{+\infty} {1 \over 2} \left(                                                                                                         
   \Phi_{2n+1/2}(x) \Phi_{2k+1/2}(x) +                                                                                                                         
   \Phi_{2n+1/2}(-x) \Phi_{2k+1/2}(-x) \right) dx                                                                                                              
$$                                                                                                                                                             
 $$                                                                                                                                                            
  = \int\limits_{-\infty}^{+\infty} { W_{2n}(x) \over (2n)!}                                                                                                   
   { W_{2k}(x) \over (2k)! } { dx \over \cosh \pi x}.                                                                                                          
 $$                                                                                                                                                            
                                                                                                                                                               
Simmilar computation for positiv integer $n$ and $k$ gives                                                                                                     
\begin{equation}                                                                                                                                               
\label{sp2}                                                                                                                                                    
   \int\limits_{-\infty}^{+\infty}                                                                                                                             
   \Phi_{-2n+1/2}(x) \Phi_{-2k+1/2}(x) dx=                                                                                                                     
   \int\limits_{-\infty}^{+\infty} { W_{2n-1}(x) \over (2n-1)!}                                                                                                
   { W_{2k-1}(x) \over (2k-1)! } { dx \over \cosh \pi x}.                                                                                                      
\end{equation}                                                                                                                                                 
                                                                                                                                                               
Functions on the left sides of (\ref{sp1}) and (\ref{sp2}) are                                                                                                 
orthonormal. Therefore for nonnegative integer $n$ and $k$,                                                                                                    
boths odd or even, we have                                                                                                                                     
\begin{equation}                                                                                                                                               
\label{spw}                                                                                                                                                    
   \int\limits_{-\infty}^{+\infty}                                                                                                                             
   {W_{n}(x) W_{k}(x) \over cosh \pi x} dx = (n!)^{2} \delta_{n,k}.                                                                                            
\end{equation}                                                                                                                                                 
If $n$ is odd and $k$ is even then the integrated function is odd                                                                                              
and the integral vanish.                                                                                                                                       
                                                                                                                                                               
                                                                                                                                                               
We may regard the set of the polynomials $W_{n}(x)$ as the system of                                                                                           
orthogonal polynomials with respect to the scalar product                                                                                                      
\begin{equation}                                                                                                                                               
\label{defsp}                                                                                                                                                  
    <f,g> \equiv  \int\limits_{-\infty}^{+\infty}                                                                                                              
    { \overline{f(x)} g(x) \over \cosh \pi x} dx.                                                                                                              
\end{equation}                                                                                                                                                 
                                                                                                                                                               
We may prove orthogonality relation (\ref{spw}) directly,                                                                                                      
not refering to Fourier transformation. We are going to use the                                                                                                
generating function $W(x,t)$.                                                                                                                                  
                                                                                                                                                               
Let us calculate in two different ways an integral                                                                                                             
\begin{equation}                                                                                                                                               
\label{ww}                                                                                                                                                     
   \int\limits_{-\infty}^{+\infty}                                                                                                                             
   {W(x,s) W(x,t) \over \cosh \pi x} dx                                                                                                                        
\end{equation}                                                                                                                                                 
On the one hand we have                                                                                                                                        
\begin{equation}                                                                                                                                               
\label{ww1}                                                                                                                                                    
   \sum\limits_{n=0}^{\infty} \sum\limits_{k=0}^{\infty}                                                                                                       
   s^{n}t^{k}  \int\limits_{-\infty}^{+\infty}                                                                                                                 
   {W_{n}(x) W_{k}(x) \over cosh \pi x} dx .                                                                                                                   
\end{equation}                                                                                                                                                 
On the other hend we have                                                                                                                                      
\begin{equation}                                                                                                                                               
\label{ww2}                                                                                                                                                    
    \int\limits_{-\infty}^{+\infty}                                                                                                                            
    { \exp ( 2x ( \arctan s + \arctan t) ) \over                                                                                                               
    \sqrt{1+s^{2}} \sqrt{1+t^{2}} \cosh \pi x } dx =                                                                                                           
    {1 \over 1 - st} = \sum\limits_{n=0}^{\infty} s^{n}t^{n}.                                                                                                  
\end{equation}                                                                                                                                                 
Comparing the coefficent of the same power $s$ and $t$ in the formulae                                                                                         
(\ref{ww1}) and (\ref{ww2}) we get (\ref{spw}). To compute the integral                                                                                        
(\ref{ww2}) we exploit the relations                                                                                                                           
\begin{equation}                                                                                                                                               
\label{r1}                                                                                                                                                     
    \int\limits_{-\infty}^{+\infty}                                                                                                                            
    {e^{2x\theta} \over \cosh \pi x} dx = {1 \over cos \theta },                                                                                               
    \quad  - { \pi \over 2 } < \theta < { \pi \over 2 }                                                                                                        
\end{equation}                                                                                                                                                 
and                                                                                                                                                            
\begin{equation}                                                                                                                                               
\label{r2}                                                                                                                                                     
    \cos (\arctan s + \arctan t) =                                                                                                                             
    { 1 -st \over \sqrt{1+s^{2}} \sqrt{1+t^{2}} }.                                                                                                             
\end{equation}                                                                                                                                                 
The integral (\ref{r1}) we can calculate using complex analysis method                                                                                         
in the very similar way as we calculated Fourier transform of the generating                                                                                   
function (\ref{ftgf}). We exploit the following property of integrated                                                                                         
function $f(x) \equiv exp( 2 x \theta ) / cos \pi x$:                                                                                                          
\begin{equation}                                                                                                                                               
\label{r3}                                                                                                                                                     
    f(x+i) = - f(x) e^{2i \theta}.                                                                                                                             
\end{equation}                                                                                                                                                 
                                                                                                                                                               
\section{Final remarks}                                                                                                                                        
{\bf 1.} Let us define the expression.                                                                                                                         
\begin{equation}                                                                                                                                               
\label{rdef}                                                                                                                                                   
    Rf(x) \equiv                                                                                                                                               
    {i \over 2} \left( {1 \over 2} -ix \right) f(x+i)-                                                                                                         
    {i \over 2} \left( {1 \over 2} +ix \right) f(x-i)                                                                                                          
    + x f(x).                                                                                                                                                  
\end{equation}                                                                                                                                                 
Then $R$ behaves as an "creation  operator"                                                                                                                    
\begin{equation}                                                                                                                                               
\label{rc}                                                                                                                                                     
   [h,R]=R,                                                                                                                                                    
\end{equation}                                                                                                                                                 
where h is defined by formula (\ref{hwn}).                                                                                                                     
Therefore                                                                                                                                                      
\begin{equation}                                                                                                                                               
\label{impl}                                                                                                                                                   
   hf = \lambda f \quad \hbox{implies} \quad                                                                                                                   
   h R f = (\lambda + 1) R f.                                                                                                                                  
\end{equation}                                                                                                                                                 
That confirms the designation "creation operator" for R.                                                                                                       
More exactly, we have                                                                                                                                          
\begin{equation}                                                                                                                                               
\label{rwn}                                                                                                                                                    
    R W_{n} = W_{n+1}.                                                                                                                                         
\end{equation}.                                                                                                                                                
                                                                                                                                                               
\noindent {\bf 2.}                                                                                                                                             
The equation                                                                                                                                                   
\begin{equation}                                                                                                                                               
\label{x1}                                                                                                                                                     
    {i \over 2} \left( {1 \over 2} -ix \right) \Phi(x+i)-                                                                                                      
    {i \over 2} \left( {1 \over 2} +ix \right) \Phi(x-i)                                                                                                       
     = {1 \over 2} \Phi(x)                                                                                                                                     
\end{equation}                                                                                                                                                 
is not only solved by the function                                                                                                                             
\begin{equation}                                                                                                                                               
\label{x2}                                                                                                                                                     
   \Phi_{1/2}(x)={2 e^{-\pi x /2} \over 1+ e^{-2 \pi x} }                                                                                                      
\end{equation}                                                                                                                                                 
but also by the function                                                                                                                                       
\begin{equation}                                                                                                                                               
\label{x3}                                                                                                                                                     
   \Phi(x) \equiv \Phi_{1/2}(x+a).                                                                                                                             
\end{equation}                                                                                                                                                 
The Fourier transform of $\Phi$ reads                                                                                                                          
\begin{equation}                                                                                                                                               
\label{x4}                                                                                                                                                     
    \Psi(p) = {1 \over \sqrt{ 2 \pi} } \int\limits_{-\infty}^{+\infty}                                                                                         
    \Phi(x) e^{-i px} dx = \Psi_{1/2}(p) e^{-ipa}                                                                                                              
\end{equation}                                                                                                                                                 
where                                                                                                                                                          
\begin{equation}                                                                                                                                               
\label{x5}                                                                                                                                                     
    \Psi_{1/2}(p)= {1+i \over 1+ie^{p} } e^{p/2}.                                                                                                              
\end{equation}                                                                                                                                                 
We have                                                                                                                                                        
\begin{equation}                                                                                                                                               
\label{x6}                                                                                                                                                     
    K \Psi(p) = \left( {1 \over 2} + a \cosh p \right) \Psi(p).                                                                                                
\end{equation}                                                                                                                                                 
and for $a \ne 0$ the function $\Psi(p)$ is not an eigenfunction of $K$.


{\bf Acknowledgement.}
We are deeply indebted to Dr. Witold Karwowski for his assistance in
formulating properly the mathematical problems in the Appendix.
We are also grateful to Dr. Andrzej Hulanicki, Dr. Jerzy Lukierski,
Dr. Helmut Reeh and Dr. Ludwik Turko for inspiring discusions.

\section*{Appendix}
Let us start with (\ref{410b}), viz.
$$
   K={i \over 2} ( \partial_{p} \cosh(p)+ \cosh(p) \partial_{p}).
\eqno{(A1)}
$$
This differential expression when applied to $C_{0}^{\infty}({\bf R})$
 defines a symmetric operator in $L^{2}({\bf R})$ as stated in
section 6. We shall denote this operator also by $K$. Indeed, if
$\Psi$ and $\Phi$ belong to $C_{0}^{\infty}({\bf R}) $  then
$$
   \int\limits_{-\infty}^{+\infty} \overline{\Psi(p)} K \Phi(p) dp
 - \int\limits_{-\infty}^{+\infty} \overline{K \Psi(p)} \Phi(p) dp
$$
$$
   = i  \int\limits_{-\infty}^{+\infty} { d \over dp }
   \left( \cosh(p) \overline{\Psi(p)} \Phi(p) \right) dp
\eqno{(A2)}
$$ 
$$
   = i \cosh(p) \overline{\Psi(p)} \Phi(p) |_{-\infty}^{+\infty} =0
$$
Let $\lambda$ belong to ${\bf C}$. Then the equation
$$
   K \Psi_{\lambda}(p) = \lambda \Psi_{\lambda}(p)
\eqno{(A3)}
$$
has the solution
$$
   \Psi_{\lambda}(p)= C_{\lambda} {1 \over \sqrt{ \cosh p} }
   \exp(-i \lambda \arctan \sinh p)
\eqno{(A4)}
$$
which is unique up to the multiplicative constant $C_{\lambda}$.
We may use this constant to normalize  $\Psi_{\lambda}$.
Clearly  $\Psi_{\lambda}$ belongs to $L^{2}({\bf R})$ as
\footnote{ For $\lambda, \mu \in {\bf R}$ and $\lambda \ne \mu$
we have
$$
  \int\limits_{-\infty}^{+\infty} \overline{\Psi_{\lambda}(p)}
  \Psi_{\mu}(p) dp =
  \overline{C_{\lambda}} C_{\mu} { \sin (\lambda - \mu) \pi / 2
  \over (\lambda - \mu) \pi /2 }.
$$
}
$$
  \int\limits_{-\infty}^{+\infty} \overline{\Psi(p)}  \Psi(p) dp =
  |C_{\lambda}|^{2}  \int\limits_{-\infty}^{+\infty} 
  { dp \over \cosh p } = |C_{\lambda}|^{2} \pi.
\eqno{(A5)}
$$

Thus the defect subspaces for the adjoint operator, $K^{+}$, are
one dimensional and the defect indices are $\{ -1 ,1 \}$. Hence $K$
has the one-parameter family of self-adjoint extensions, which can be
defined as follows. Let
$$
   \alpha \in {\bf C}, \quad |\alpha|=1
\eqno{(A6)}
$$
and define
$$
  M_{\alpha}=\{ \Phi \in C^{\infty}({\bf R}):
  \lim_{p \rightarrow - \infty} \left( \sqrt{\cosh p} \; \Phi(p) \right) =
  \alpha \lim_{p \rightarrow + \infty} \left( \sqrt{\cosh p} \;
  \Phi(p)\right) \}.
\eqno{(A7)}
$$
Let us further define the operator $K_{\alpha}$ as
$$
   K_{\alpha} \Phi(p) = {i \over 2}
   ( \sinh( p) + 2 \cosh(p) {d \over dp}) \Phi(p), \quad
   \Phi \in M_{\alpha}.
\eqno{(A8)}
$$
The operator $K_{\alpha}$ is essentially self-adjoint.
To see this we observe that
$$
   Ran(K_{\alpha} \pm i \jjed )
$$
is dense in $L^{2}({\bf R})$. Indeed, taking
$$
   \Psi=\Psi_{\pm i},
$$
we see that (A2) does not hold for all $\Phi \in M_{\alpha}$.
Thus $\Psi$ do not belong to the domain
$   D(K^{+}_{\alpha})  $
of $K^{+}_{\alpha}$. Consequently if $\Psi \in D(K^{+}_{\alpha}) $
then
$$
   K \Psi = K^{+}_{\alpha} \Psi = \pm i \Psi
   \quad  \hbox{implies} \quad \Psi =0.
\eqno{(A9)}
$$
This implies that if
$$
   (\Psi, (K_{\alpha} \mp i \jjed )\Phi )=
   ( (K^{+}_{\alpha} \pm i \jjed ) \Psi, \Phi ) =0
\eqno{(A10)}
$$
for all $\Phi \in M_{\alpha}$ then $\Psi=0$ and hence
$$
      Ran( K_{\alpha} \pm i \jjed )
\eqno{(A11)}
$$
is dense and $K_{\alpha}$ is essentially self-adjoint. We shall denote the
closure of $K_{\alpha}$ by $\overline{ K_{\alpha} }$.

Let us now consider the function (A4).
For real $\lambda$ we have
$$
  \lim_{p \rightarrow - \infty}
  \left( \sqrt{\cosh p} \; \Psi_{2n+\gamma}(p) \right) =
  \alpha \lim_{q \rightarrow + \infty}
  \left( \sqrt{\cosh p} \; \Psi_{2n+\gamma}(p) \right)
\eqno{(A12)}
$$
where
$$
   \alpha=e^{i \pi \gamma }.
$$
We conclude that for any $0 \leq \gamma < 2$ the functions
$$
    \{ \Psi_{2n + \gamma } \}_{ n \in {\bf Z}}
\eqno{(A13)}
$$
form a system of eigenvectors for the self-adjoint operator
$$
\overline{K_{\alpha}}, \quad \alpha=e^{-i\gamma \pi}.
\eqno{(A14)}
$$
The corresponding eigenvalues are
$$
   \{ 2n+\gamma \}_{ n \in {\bf Z}}
\eqno{(A15)}
$$

It follows from (A5) that if $C_{2n+\gamma}=1/\sqrt{\pi}$ then
$\Psi_{2n+\gamma}$ are normalized. We show that for any
$$
  0 \leq \gamma < 2
$$
the system
$$
  \{ \Psi_{2n+\gamma} \}_{ n \in {\bf Z}}
$$
is an orthonormal basis for $L^{2}({\bf R})$. It is sufficient to show that
for any
$$
   f \in L^{2}({\bf R})
$$
there is
$$
    \sum\limits_{n \in {\bf Z} } |(f,\Psi_{2n+\gamma})|^{2}=||f||^{2}
\eqno{(A16)}
$$
where $(.,.)$ and $||.||$ denote the inner product and the norm
in $L^{2}$ resp. To prove (A16) let us observe that
$$
   (f,\Psi_{2n+\gamma})= {1 \over \sqrt{\pi}}
   \int\limits_{-\infty}^{+\infty}
   \overline{f(p)} (\cosh p)^{-1/2} \exp(-i(2n+\gamma) \arctan \sinh p) dp.
\eqno{(A17)}
$$
Let us change the variables as follows
$$
    \sinh p=\tan { \theta \over  2},\quad -\pi \leq \theta \leq \pi,
\eqno{(A18)}
$$
 $$
    \cosh p = {1 \over \cos { \theta \over 2} }, \quad
    dp = { d \theta \over \cos { \theta \over 2 } }.
 $$
Then (A17) takes the form
$$
  (f,\Psi_{2n+\gamma})=
  {1 \over 2 \sqrt{\pi} } \int\limits_{-\pi}^{+\pi}
  \overline{ f ( \hbox{arcsinh}\;  \tan { \theta \over 2 } ) }
  (\cos { \theta \over 2 } )^{- { 1 \over 2 } }
  \exp({ -i\gamma \theta \over 2}) \exp(-i n \theta) d \theta.
\eqno{(A19)}
$$
We get further
$$
   \sum\limits_{n \in {\bf Z}} |(f,\Psi_{2n+\gamma})|^{2}=
   {1 \over 4 \pi} \sum\limits_{n \in {\bf Z}}
   |\int\limits_{-\pi}^{\pi} \overline{ \tilde{f} (\theta) } 
   e^{-in\theta} d\theta |^{2}=
$$
$$
= {1\over 2} \int\limits_{-\pi}^{\pi} |\tilde{f}(\theta)|^{2} d\theta
=\int\limits_{-\infty}^{+\infty} |f(q)|^{2} dq = ||f||^{2}
\eqno{(A20)}
$$
where
$$
   \tilde{f} (\theta)\equiv 
   f( \hbox{arcsinh}\; \tan {\theta \over 2} )
   (\cos {\theta \over 2} )^{-1/2} e^{i\gamma \theta/2}
\eqno{(A21)}
$$
Formula (A20) is the Parseval equality.

\end{document}